\documentclass[aps,prb,twocolumn,amsmath,amssymb,nofootinbib,superscriptaddress,floatfix,eqsecnum]{revtex4-1}

\usepackage{amsmath}
\usepackage{amssymb}
\usepackage{amsthm}
\usepackage[dvipdf]{color}
\usepackage{graphicx}
\usepackage{epsfig}
\usepackage{epstopdf}
\usepackage{dcolumn}
\usepackage{bm}

\begin{document}
\title{Random Nonlinear Infinite-Level-System Model for Amorphous Solid Phonon Echo and Saturation Phenomena}

 \author{Di Zhou$^{1}$, Anthony J. Leggett}

 \affiliation{
 Department of Physics,
  University of Illinois at Urbana-Champaign,
  1110 West Green St, Urbana,
   Illinois 61801, USA
 }

\date{\today}

\begin{abstract}
The first two successful predictions for amorphous solid experiments by tunneling-two-level system (TTLS) was phonon echo\cite{Golding1976} and saturation\cite{Graebner1983} phenomena. In this paper by generalizing TTLS to infinite-level-system model with certain randomness and nonlinearity, we prove that phonon echo is a stimulated emission process. The classical infinite-level-system (taking $\hbar\to 0$ limit) cannot find saturation effect, while quatum infinite-level-system can find it with certain nonlinearity assumed. We also prove that without randomness or nonlinearity neither phonon echo nor saturation exists in arbitrary infinite-level-system.
\end{abstract}

\maketitle

\section{Introduction}
Glass display a number of universal properties which cannot be found in crystalline solids, not only below the temperatures of 1K, but also between 1K and 30K\cite{Pohl2002} . These properties were first well studied by Zeller and Pohl\cite{Zeller1971} like low temperature heat capacity and thermal conductivity. In 1972, Anderson, Halperin and Varma group\cite{Anderson1972} and Phillips\cite{Phillips1987} independently set up a model which was later named tunneling-two-level system (TTLS). Together with a certain random distribution the parameter of it\cite{Hunklinger1986}, this model not only successfully explained existing experiments such as linear temperature dependence of heat capacity, but also predict new experiments such as phonon echo\cite{Golding1976} and saturation\cite{Graebner1983} phenomena.

The original experiment of phonon echo phenomena is to plug in two short ultrasonic pulses with identical frequency $f\sim 0.68$GHz\cite{Golding1976} and a time delay between them $\tau\sim0.3\times 10^{-6}$s. After the reflection of sample's boundary, we observe two reflecting signals. However, after the second pulse with a time delay of $\tau$ we observe a third one with the same frequency $f$. This spontaneously generated pulse which was later referred to as ``phonon echo" cannot be observed in crystalline solids. Phonon echo experiment is similar to spin echo\cite{Hahn1950}, in which we plug in two magnetic pulses to precess a set of spins. Such a spins system will spontaneously generate a huge magnetic signal following the second input magnetic pulse with the magnitude comparable to it. Phonon echo can be automatically explained within TTLS model by making analogy to spin echo effect, because both of two-level-system Hamiltonian and the coupling between two-level-system and phonon strain field can be represented by Pauli matrices. D. A. Parshin and his collaborators\cite{Parshin2009, Parshin2009April, Parshin2010} has derived a complete theory beyond two-level-system, that for an arbitrary multiple-level-system with randomness and anharmonicity, one can always obtain phonon echo phenomena. In this paper by setting up the coupling btween stress tensor and phonon strain field, we prove that phonon echo essentially comes from off-diagonal matrix elements of stress tensor operators in the basis of glass non-elastic part of Hamiltonian. In addition, we prove that the mechnism when phonon echo occurs is stimulated phonon emission process rather than spontaneous emission process which explains why echo signal is comparable to input ultrasonic pulses. 

Saturation experiment\cite{Graebner1983} is to plug in two phonon pulses. Unlike phonon echo, the first ``saturating pulse" is much stronger than the second ``testing pulse". The second testing pulse is plugged in with a time delay much smaller than thermal relaxation time of TTLS. Since the energy level structure of amorphous solid is the summation of long-wavelength phonon modes and a set of TTLS, putting in a strong ``saturating pulse" will saturate these two-level-systems, resulting in the large decrease of attenuation $\alpha=l^{-1}$, implying that the phonon absorption ability is saturated by the first strong pulse. Within TTLS model energy eigenstates can go no more than first excitation so that saturation effect is automaticallly proved. In this paper we demonstrate even for infinite-level-system we can still find saturation effect.

The paper is organized as follows: in section 2 we develop our infinite-level-system by expanding amorphous non-elastic Hamiltonian in orders of strain field $e_{ij}$; in section 3 we prove that phonon echo only occurs when stimulated emission process happens. We compare different cases and demonstrate randomness and nonlinearity are necessary for phonon echo, while for an arbitrary system without randomness or nonlinearity phonon echo never exists; in section 4 we prove not only for finite-level-system but also for quantum infinite-level-system, phonon saturation can be found with the presence of nonlinearity, while for classical infinite-level-system (with $\hbar\to 0$ limit) we can never find saturation phenomena. In conclusion phonon saturation is a result of quantum physics. With these results we can roughly describe the energy structure of amorphous solid Hamiltonian.

\section{The Set up of Infinite-Level-System}
Due to TTLS theory, the amorphous solid Hamiltonian is the summation of elastic (phonon) Hamiltonian, a set of tunneling-two-level-systems and phonon-TTLS couplings\cite{Hunklinger1986}. Because we are interested in TTLS, usually we omit phonon part Hamiltonian. In the basis when TTLS is diagonalized, the Hamiltonian for amorphous solid is written as 
\begin{eqnarray}\label{1}
\hat{H} & = & \int d^3 x\,\mathcal{H}(\vec x)\nonumber \\
\mathcal{H}(\vec x) & = & \frac{1}{2}E(\vec x)\sigma_3\nonumber \\
 & + & \frac{\gamma(\vec x)}{2}e \left(D(\vec x)\sigma_3+M(\vec x)\sigma_1\right)
\sin(\omega t-k\cdot x+\theta)\nonumber \\
\end{eqnarray}
where $e$ is phonon strain; $\sigma_{1,2,3}$ are Pauli matrix representations; $E(\vec x)=\sqrt{\Delta(\vec x)^2+\Delta_0(\vec x)^2}$ is the TTLS energy splitting, where $\Delta(\vec x)$ and $\Delta_0(\vec x)$ are TTLS diagonal and off-diagonal matrix elements at position $\vec x$. For different positions $\vec x$ they are randomly distributed; $D(\vec x)=\Delta(\vec x)/E(\vec x)$ and $M(\vec x)=\Delta_0(\vec x)/E(\vec x)$ are diagonal and off-diagonal matrix elements of TTLS-phonon coupling; $\gamma(\vec x)$ is the coupling constant between TTLS and phonon strain which is also randomly distributed over space $\vec x$. Since the Hamiltonian is written in the representation of spin operators, one can easily prove ``phonon echo" by making analogy to spin echo effect which was first observed by Hahn\cite{Hahn1950}. On the other hand TTLS can go no higher than first excitation, J$\ddot{\rm a}$ckle\cite{{Jackle1972}} pointed out that phonon energy absorption per unit time is maximized when TTLS ground state and first excitation population are equally distributed with $P_0=P_1=1/2$. Because for an arbitrary two-level-system with certain randomness in their matrix elements echo effect always exists, and for an arbitrary finite-level-system one can always prove saturation phenomena, in this paper we are interested in proving them in arbitrary infinite-level-system.

Let us consider a block of amorphous solid with the dimension $L$ much greater than atomic distance $a\sim 10\AA$. The elastic strain $e_{ij}(\vec x, t)$ is defined as the spacial derivative of displacement $\vec u(\vec x, t)$ for the matter located at $\vec x$: 
\begin{eqnarray}\label{2}
e_{ij}(\vec x, t)=\frac{1}{2}\left(\frac{\partial u_i(\vec x, t)}{\partial x_j}+\frac{\partial u_j(\vec x, t)}{\partial x_i}\right)
\end{eqnarray}
By subtracting elastic part of amorphous Hamiltonian, we write $\hat{H}$ to stand for the non-elastic part Hamiltonian of it. We expand it in orders of elastic strain $e_{ij}(\vec x, t)$ in long wavelength limit $L\gg\lambda\gg a$:
\begin{eqnarray}\label{3}
\hat{H}(t)=\int d^3 x\,\left(\hat{\mathcal{H}}_0(\vec x)+\sum_{ij}e_{ij}(\vec x, t)\hat{T}_{ij}(\vec x)\right)+\mathcal{O}(e_{ij}^2)\nonumber \\
\end{eqnarray}
where the definition of stress tensor $\hat{T}_{ij}(\vec x)$ is the first order derivative of Hamiltonian with respect to strain field
\begin{eqnarray}\label{4}
\hat{T}_{ij}(\vec x)=\frac{\delta \hat{H}(t)}{\delta e_{ij}(\vec x, t)}
\end{eqnarray}
The matrices $\hat{T}_{ij}(\vec x)$ and $\hat{\mathcal{H}}_0(\vec x)$ are generalization of TTLS operators. Since the stress-strain interacting term $e_{ij}\hat{T}_{ij}$ contains phonon field $e_{ij}$, the exchange of virtual phonons will give an effective RKKY-type interaction via stress tensor products\cite{Joffrin1976}:
\begin{eqnarray}\label{13}
\hat{V}=\int d^3xd^3x'\sum_{ijkl}\Lambda_{ijkl}(\vec x-\vec x')\hat{T}_{ij}(\vec x)\hat{T}_{kl}(\vec x')
\end{eqnarray}
where the coefficient $\Lambda_{ijkl}(\vec x-\vec x')$ was first derived by Joffrin and Levelut\cite{Joffrin1976}. A further detailed correction to this coefficient was given by D. Zhou and A. J. Leggett\cite{Zhou2015-1}:
\begin{eqnarray}\label{14}
{\Lambda}_{ijkl}(\vec x-\vec x') = -\frac{\tilde{\Lambda}_{ijkl}(\vec n)}{8\pi\rho c_t^2|\vec x-\vec x'|^3}
\end{eqnarray}
\begin{eqnarray}\label{15}
\tilde{\Lambda}_{ijkl} & = & \frac{1}{4}\bigg\{(\delta_{jl}-3n_jn_l)\delta_{ik}+(\delta_{jk}-3n_jn_k)\delta_{il}\nonumber \\
 & {} & +(\delta_{ik}-3n_in_k)\delta_{jl}
+(\delta_{il}-3n_in_l)\delta_{jk}\bigg\}\nonumber \\
 & {} & +\frac{1}{2}\left(1-\frac{c_t^2}{c_l^2}\right)\bigg\{-(\delta_{ij}\delta_{kl}+\delta_{ik}\delta_{jl}+\delta_{jk}\delta_{il})\nonumber \\
 & {} & +3(n_in_j\delta_{kl}+n_in_k\delta_{jl}+n_in_l\delta_{jk}\nonumber \\
 & {} & +n_jn_k\delta_{il}+n_jn_l\delta_{ik}+n_kn_l\delta_{ij})-15n_in_jn_kn_l\bigg\}\nonumber \\
\end{eqnarray}
where $\vec n$ is the unit vector of $\vec x-\vec x'$, and $i,j,k,l$ runs over $1,2,3$ cartesian coordinates. We name Eq.(\ref{13}) non-elastic stress-stress interaction. Finally, for dielectric amorphous solids, electric dipole-dipole interaction should be taken into account as well: $\hat{V}_{\rm dipole}=\int d^3xd^3x'\sum_{ij}\mu_{ij}(\vec x-\vec x')\hat{p}_{i}(\vec x)\hat{p}_{j}(\vec x')$ where $\mu_{ij}(\vec x-\vec x')=(\delta_{ij}-3n_in_j)/8\pi\epsilon|\vec x-\vec x'|^3$. In the rest of this paper we always use the approximation to replace $\vec x-\vec x'$ by $\vec x_{s}-\vec x_{s'}$ for the pair of the $s$-th and $s'$-th blocks, when $\vec x_{s}$ denotes the center of the $s$-th block, and that $\int_{V^{(s)}}\hat{T}_{ij}(\vec x)d^3x=\hat{T}_{ij}^{(s)}$ is the uniform stress tensor of the $s$-th block. Also, from now on we use $e_{ij}^{(s)}(t)$ to denote the phonon strain field $e_{ij}(\vec x, t)$ located at the $s$-th block. By chopping a $L\times L\times L$ amorphous block into identical unit blocks with the dimension $L_0\times L_0\times L_0$, we get these $(L/L_0)^3$ unit blocks' total Hamiltonian without external strain field 
\begin{eqnarray}\label{16}
\hat{H} & = & \sum_{s}^{(L/L_0)^3}\hat{H}_0^{(s)}\nonumber \\
 & + & \sum_{s\neq s'}^{(L/L_0)^3}\left(\sum_{ijkl}\Lambda_{ijkl}^{(ss')}\hat{T}_{ij}^{(s)}\hat{T}_{kl}^{(s')}+\sum_{ij}\mu_{ij}^{(ss')}\hat{p}_i^{(s)}\hat{p}_j^{(s')}\right)\nonumber \\
\end{eqnarray}

The many body interaction $\hat{V}+\hat{V}_{\rm dipole}$ may complicate echo and saturation problems: a ``Heiserberg"-like behavior might be taken into consideration, leading to the decohenrence of echoing phonons' phases. However, due to A. Morello, P. C. E. Stamp and I. S. Tupitsyn's argument\cite{Stamp} that as long as the many body interaction $\hat{V}+\hat{V}_{\rm dipole}\ll$ the off-diagonal matrix elements of multiple-level-system Hamiltonian $\hat{H}_0^{(s)}$, many body interaction decoherence will be small enough not to affect multiple-level-systems' wavefunction phase coherence. To estimate the many body interaction strength of multiple-level-system versus the off-diagonal matrix elements of it, we use TTLS parameters' order of magnitude as follows: 

(1) To contribute to thermal properties of amorphous solids around $T=1$K, off-diagonal matrix elements of $\hat{H}_0^{(s)}$ have to be of order $1{\rm K}\sim 10^{-4}$eV; 

(2) Dielectric amorphous materials such as BK7 and SiO$_2$ have greatest electric dipole moments. Their electric dipole interaction strength $\hat{V}_{\rm dipole}$ are of the same order of magnitude as non-elastic stress-stress interaction $\hat{V}$, see S. Hunklinger and M. V. Schickfus' measurements\cite{Schickfus1981}. The strength of $\hat{V}$ and $\hat{V}_{\rm dipole}$ are of order\cite{Stamp} $0.127{\rm K}\sim 10^{-5}{\rm eV}$, one order smaller than $\hat{H}_0^{(s)}$ off-diagonal matrix elements $10^{-4}$eV. Other amorphous solids' electric dipole interactions are even smaller. 

In conclusion, many body interaction $\hat{V}+\hat{V}_{\rm diople}$ decoherence is not strong enough to affect the wavefunction coherence of mutiple-level-system. In the rest of this paper we neglect many body interaction and focus on Hamiltonian $
\hat{H}=\sum_{s}^{(L/L_0)^3}\left(\hat{H}_0^{(s)}+e_{ij}^{(s)}(t)\hat{T}_{ij}^{(s)}+\mathcal{O}(e^2)\right)$.

We define $|n^{(s)}\rangle$ and $E_n^{(s)}$ to be the $n$-th level eigenstate and eigenvalue for Hamiltonian $\hat{{H}}_0^{(s)}$. Generally speaking, $\hat{T}_{ij}^{(s)}$ is not diagonal in the basis of $|n^{(s)}\rangle$. In the rest of this paper we write $\hat{{H}}_0$, $\hat{T}_{ij}$, $e_{ij}(t)$ and $|n\rangle$ to stand for $\hat{{H}}_0^{(s)}$, $\hat{T}_{ij}^{(s)}$, $e_{ij}^{(s)}(t)$ and $|n^{(s)}\rangle$ for convenience. Within the first order expansion in $e_{ij}(t)$, the amorphous solid Hamiltonian is free part $\hat{{H}}_0$ perturbed by time-dependent term $e_{ij}(t)\hat{T}_{ij}$.

\section{Microscopic Explaination for Phonon Echo}
The original phonon echo experiment by Golding and Graebner\cite{Golding1976} has several time scales: (1) tunneling-two-level-system relaxation time $T_1$, which is now the relaxation time for a certain pair of energy levels $|n\rangle$ and $|m\rangle$ in resonance with phonon pulse $\hbar\omega=E_m-E_n$: $T_1\approx 2\times 10^{-4}$s at temperature $T=20$mK. We should note $T=20$mK is the starting temperature of ``linear temperature dependence of heat capacity of amorphous solid" \cite{Hunklinger1986}, which implies the TTLS (infinite-level-systems) have taken part in phonon echo process; (2) characteristic thermal phonon frequency $\omega_T$, obtained by $\hbar\omega_T=k_BT\Rightarrow \omega_T\sim 2$GHz, the period is of order $3\times 10^{-9}$s; (3) input ultrasonic phonon pulse frequency $\omega=0.68$GHz with exponential decay rate $\tau_0=10^{-7}$s. The frequency is roughly the same order of magnitude as thermal phonon frequency; (4) the time delay between second and first phonon pulses: $\tau=0.3\times 10^{-6}$s. 

These time scales in orders, thermal relaxation time $T_1\gg $ time delay between two pulses $\tau\gg $ pulse decay rate $\tau_0 \gg $ pulse period $2\pi/\omega$. For convenience we plot their relation in the next page. Thermal relaxation time is much greater than the time scale of phonon echo process so we don't need to consider it; two-pulse time delay $\tau$ is the next greatest time scale, which means when the second pulse is applied, the first (exponentially decay) pulse has already faded away; pulse decay rate $\tau_0$ is much greater than ultrasonic period $2\pi/\omega$, which means the external two pulses oscillate enough times (roughly $\sim 100$ times) before they decay, therefore Fermi golden rule is valid in our problem. Neverthless, pulse decay rate $\tau_0$ brodens the absorption frequency bandwidth so that the previous absorption function $\delta(E_m-E_n-\hbar\omega)$ becomes a lorentzian form $1/[(\omega_{m}-\omega_n-\omega)^2+\hbar^2/\tau_0^{2}]$. This will play an important role in the realization of phonon echo effect. 
\begin{figure}[hp]
\includegraphics[scale=0.43]{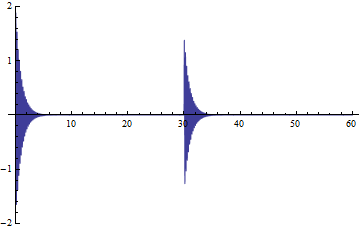}
\caption{Two phonon pulses with time separation $\tau\gg$ pulse decay rate $\tau_0\gg$ pulse oscillation period $2\pi/\omega$; these two pulses have identical frequency $\omega$, while their amplitude and decay rate $\tau_0$ are not necessarily the same. }  
\end{figure}

Using the above data analysis let's solve time-dependent Schrodin$\ddot{\rm g}$er equation and calculate the transition rate $C_{mn}$ between $n$-th and $m$-th level for $\hat{T}_{ij}$:
$
\left(\hat{{H}}_0+\sum_{ij}e_{ij}(t)\hat{T}_{ij}\right)\psi(t)=i\hbar\frac{\partial }{\partial t}\psi(t)
$. We get 
\begin{eqnarray}\label{5}
\lim_{t\to\infty}C_{mn}(t) & = & \frac{1}{i\hbar}\int_0^{\infty}e^{i\omega_{mn}t'}\sum_{ij}e_{ij}(t')\langle m|\hat{T}_{ij}|n\rangle dt'\nonumber \\
 & = & \frac{e^{i\theta}}{\frac{i\hbar}{\tau_0}+(\omega_{mn}-\omega)}\sum_{ij}e_{ij}\langle m|\hat{T}_{ij}|n\rangle
\end{eqnarray}
where $\omega_{mn}=\omega_m-\omega_n=(E_m-E_n)/\hbar$, the energy spacing between $n$-th and $m$-th levels; $\theta$ is the starting phase of phonon pulse. In later discussions we always use $C_{mn}$ and $C_{nm}$ to represent the transition rate between $m$-th and $n$-th levels. In experiments there could be more than one pair of levels in resonance with $\omega$, however the calculation and analysis are the same as single-pair resonance case.

We choose the first pulse plugging in time to be $t=0$. At $t<0$, infinite-level-systems are in thermal equilibrium with temperature $T=20$mK. Therefore for the $n$-th eigenstate of infinite-level-system, the wavefunction is 
$
|n, t\rangle = a_ne^{i\omega_nt+i\theta_n}|n\rangle
$, where $\theta_n$ is the initial thermal random phase and $\omega_n=E_n/\hbar$ is the energy eigenvalue. At $t=0$ the first ultrasonic pulse is plugged in with frequency $\omega$ and decay rate $\tau_0$. After the first pulse is fully applied (for example, at time $t>5\tau_0$) the wavefunction for $n$-th and $m$-th levels are given by 
\begin{eqnarray}\label{6}
|n, t\rangle & = & \left[a_n(1-C^{(1)}_{nm})e^{i\theta_n}+a_mC^{(1)}_{mn}e^{i\theta_m}\right]e^{i\omega_n t}|n\rangle \nonumber \\
|m, t\rangle & = & \left[a_m(1-C^{(1)}_{mn})e^{i\theta_m}+a_nC^{(1)}_{nm}e^{i\theta_n}\right]e^{i\omega_m t}|m\rangle
\nonumber \\
\end{eqnarray}
where $C_{nm}^{(1)}$ is the transition rate from $n$-th level to $m$-th level due to the first ultrasonic pulse. At $t=\tau$ a second pulse is plugged in. After the second pulse is fully applied, $t>\tau+5\tau_0$ the wavefunction for $n,m$-th levels are 
\begin{eqnarray}\label{7}
 & {} & |n, t\rangle\nonumber \\
 &  = & \left[a_n(1-C^{(1)}_{nm})e^{i\theta_n}+a_mC^{(1)}_{mn}e^{i\theta_m}\right](1-C_{nm}^{(2)})e^{i\omega_n t}|n\rangle \nonumber \\
 & + & \left[a_m(1-C^{(1)}_{mn})e^{i\theta_m}+a_nC^{(1)}_{nm}e^{i\theta_n}\right]C_{mn}^{(2)}e^{i\omega_m \tau+i\omega_n (t-\tau)}|n\rangle\nonumber \\
\nonumber \\
 & {} & |m, t\rangle\nonumber \\
 & = & 
\left[a_m(1-C^{(1)}_{mn})e^{i\theta_m}+a_nC^{(1)}_{nm}e^{i\theta_n}\right](1-C^{(2)}_{mn})e^{i\omega_m t}|m\rangle \nonumber \\
 & + & \left[a_n(1-C^{(1)}_{nm})e^{i\theta_n}+a_mC^{(1)}_{mn}e^{i\theta_m}\right]C_{nm}^{(2)}e^{i\omega_n \tau+i\omega_m (t-\tau)}|m\rangle\nonumber \\
\end{eqnarray}
For other eigenstates not in resonance with external pulse, their phases with quantum number $p$ are $\Theta_p=\omega_pt+\theta_p$. Throughout the entire experiment, infinite-level-system eigenstates spontaneously emit phonons under the influence of perturbation $e_{ij}(t)\hat{T}_{ij}$. However since thermal relaxation time is much greater than any other time scale we can hardly observe any spontaneous emission signal. From the previous Eq.(\ref{7}), the $m$-th state wavefunction has four components, their phases $\Theta_m^*$ are $\Theta_m^*=\theta_m+\omega_mt$, $\theta_n+\omega_mt$, $\theta_n+\omega_n\tau+\omega_m(t-\tau)$ and $\theta_m+\omega_n\tau+\omega_m(t-\tau)$. Similarly the $n$-th state wavefunction has four components with their phases $\Theta_n^*=\theta_n+\omega_nt$, $\theta_m+\omega_nt$, $\theta_m+\omega_m\tau+\omega_n(t-\tau)$ and $\theta_n+\omega_m\tau+\omega_n(t-\tau)$. Please note $\Theta_{m,n}^*$ are random numbers because $|m\rangle, |n\rangle$ energy eigenvalues $\omega_m, \omega_n$ and their wavefunction thermal equilibrium starting phases $\theta_m, \theta_n$ are random numbers: for the infinite-level-system at position $\vec x$, $\omega_{m}, \omega_n, \theta_m, \theta_n$ are different for different $m, n$ level pairs in resonance wth ultrasonic frequency $\omega$; on the other hand $\omega_{m}, \omega_n, \theta_m, \theta_n$ are random functions of spacial coordinate $\vec x$. What is more, recall the pulse decay rate $\tau_0^{-1}$, resonance absorption spectrum is broadened from delta-function to lorentzian form $1/((\omega_m-\omega_n-\omega)^2+\hbar^2/\tau_0^2)$ so that $(\omega_{m}-\omega_n)$ has random deviation from $\omega$ of the order $\pm \hbar/\tau_0$.

We consider several possibilities regarding phonon emission process: (1) a $p$-th state with phase $\Theta_p=\omega_pt+\theta_p$ emits a phonon to relax back to state $q$ with phase $\Theta_q=\omega_qt+\theta_q$ ($q\neq m, n; q<p$), the emitting phonon phase is $\Theta_p-\Theta_q$; (2) state $p$ relaxes back to state $m$ or $n$ with phase $\Theta_m$ or $\Theta_n$, where $\Theta_{m,n}$ not necessarily equal to $\Theta_{m,n}^*$, the emitting phonon phase is $\Theta_p-\Theta_{m,n}$; (3) state $m$ or $n$ relaxes back to state $q$, the emitting phonon phase is $\Theta_{m,n}^*-\Theta_q$; (4) state $m$ relaxes back to state $n$ with phase $\Theta_n$ not necessarily equal to $\Theta_n^*$, the emitting phonon phase is $\Theta_m^*-\Theta_n$. Still, the above quantities $\Theta_{p,q,m,n}$ are random. 

Together with the randomness of $\Theta_{m,n}^*$ we know all these four quantities $\Theta_p-\Theta_q$, $\Theta_p-\Theta_{m,n}$, $\Theta_{m,n}^*-\Theta_q$ and $\Theta_m^*-\Theta_n$ are random numbers, which means corresponding spontaneously emitting phonons are incoherent at arbitrary time before $2\tau$. Not only the relaxation time $T_1\sim 2\times 10^{-4}$s is greater more than 2 orders of magnitude than any other times scales, so the spontaneous emission signal is negligible compared to input pulses, but also the phonon phase cancel each other to further weaken spontaneous signal.

However, at $t=2\tau$ a strong coherent signal emerges. If one $m$-th state accidentally emits phonon and relaxes back to state $n$ with phase $\Theta_n=\Theta_n^*$, then the emitting phonon phase $\Theta_m^*-\Theta_n^*$ has 9 possibilities: $\pm(\theta_m-\theta_n)+(\omega_m-\omega_n)t$, $\pm(\theta_m-\theta_n)+(\omega_m-\omega_n)(t-\tau)$, $\pm(\theta_m-\theta_n)+(\omega_m-\omega_n)(t-2\tau)$, $(\omega_m-\omega_n)t$, $(\omega_m-\omega_n)(t-\tau)$, $(\omega_m-\omega_n)(t-2\tau)$. Although they are still random at any time $\tau+5t_0<t<2\tau$, one of them turns out to be not random at $t= 2\tau$: 
\begin{eqnarray}\label{8}
 t=2\tau: (\omega_m-\omega_n)(t-2\tau)=0\quad\quad t>\tau+5t_0 
\end{eqnarray}
One may also notice at $t=0$ and $\tau$, $(\omega_m-\omega_n)t=0$ and $(\omega_m-\omega_n)(t-\tau)=0$ are not random as well, however the wavefunction in Eq.(\ref{7}) is written after the second pulse applied, i.e., $t>\tau+5t_0$, so the solutions $t=0, \tau$ are fake; from Eq.(\ref{6}) we also demonstrate no coherent phonon generated during $\tau>t>0$. At $t=2\tau$ phonons emitted from different level pairs $m,n$ and different positions $\vec x$ have the identical phase $\Theta_m^*-\Theta_n^*=0$, resulting in the direct summation over their amplitudes. In experiment at $t=2\tau$ we observe a strong echo signal spontaneously generated by amorphous sample.

It seems echo phenomena has been explained within spontaneous emission process. However the fact that echo amplitude is comparable to input pulse leads to several questions: (1) state $m$ can not only go back to state $n$, but also relax to other states $q<m$; (2) the choice of state $n$ phase $\Theta_n$ is arbitrary, only a small fraction of $\Theta_n$ equals to $\Theta_n^*$; (3) thermal relaxation time $T_1$ is so huge that spontaneous emission is still negligible, regardless of phonon phase coherence. With these considerations we demonstrate instead of spontaneous emission process, echo signal comes from stimulated emission process which makes the amplitude comparable to input pulses. At $t=2\tau$, as long as one phonon emitted from $m,n$ levels with frequency $\omega-\tau_0^{-1}<\omega_{mn}<\omega+\tau_0^{-1}$ and phase $\Theta_m^*-\Theta_n^*=0$ is occasionally emitted, other energy level pairs $m', n'$ next to it with $\omega-\tau_0^{-1}<\omega_{m'n'}<\omega+\tau_0^{-1}$ get stimulated and emit phonons with the same phase and roughly the same frequency within deviation $\pm\tau_0^{-1}$. The importance of pulse decay rate $\tau_0$ is that it largely increases stimulated emission chance. The more phonons emitted, the more energy level pairs get stimulated. Eventually the entire set of level pairs emit phonons simultaneously all over the space. Echo signal amplitude is therefore comparable to input pulses.

From the above discussions we see why the energy level randomness over spacial coordinate $\vec x^{(s)}$ of $\hat{{H}}_0^{(s)}$, and the energy level nonlinearity (energy levels are not equally spaced) play the central role in echo phenomena. Suppose we have a Hamiltonian $\hat{{H}}_0^{(s)}+\sum_{ij}e_{ij}^{(s)}(t)\hat{T}_{ij}^{(s)}$ with nonlinearity but without randomness , and another Hamiltonian with randomness but without nonlinearity. First, the system without randomness has a set of infinite-level-systems with identical energy level structures. When the first pulse is plugged in all energy level pairs $m, n$ in resonance with ultrasonic frequency $\omega$ are excited beyond thermal equilibrium, with the same frequencies $\omega_m, \omega_n$ over space $\vec x^{(s)}$. Right after the first pulse as long as one phonon is spontaneously emitted, the entire $m, n$ level pairs get stimulated instantly. No phonon energy stored in infinite-level-system and echo cannot be observed. Second, for a system without nonlinearity in level structure, i.e., the energy levels of $\hat{{H}}_0^{(s)}$ are equally spaced, if we plug in an ultrasonic pulse either the entire energy levels absorb phonons, or none of them can. Only a set of input frequencies with $\omega=n\Omega$ can be absorbed, where $\Omega$ is the energy spacing of infinite-level-system and $n$ is integer. The absorption spectrum width is so narrow (of order $\pm\hbar/\tau_0$) that experimentally it is almost unrealizable. Even if infinite-level-system successfully absorb phonons, phonon echo still canot exist since as long as one phonon is spontaneously emitted from a certain pair of levels, other level pairs will get stimulated simultaneously. Finally, it is obvious that echo effect does not exist in any system without randomness and nonlinearity.

\section{Microscopic Explaination for Phonon Saturation}
Phonon saturation is obvious for arbitrary finite-level-system because the energy levels are capped. In this section our purpose is to prove saturation in infinite-level-system with increasing energy spacing. It is impossible to prove saturation effect with decreasing energy spacing, for example, Hydrogen atom with energy level $E_n=-13.6{\rm eV}/n^2$ can be ionized if the imcoming photon energy exceeds $13.6$eV.

First of all, in classical nonlinear oscillator we can not find saturation phenomena. The simplest case is one-dimensional Duffing oscillator, other kinds of unbounded nonlinear oscillator show the same behavior.
\begin{eqnarray}\label{9}
{H}(t)=\frac{p^2}{2m}+\frac{1}{2}m\omega^2 x^2 +\frac{1}{4}\gamma x^4+F(t)x
\end{eqnarray}
where $\gamma>0$ is the quatic nonlinearity coefficient. In 1918 it was Georg Duffing\cite{Duffing1918} who first got the analytical solution for the homogeneous classical equation of motion of Eq.(\ref{9}) when external force $F(t)=0$. However, instead of a closed form, a chaotic solution\cite{Nagy} emerges for inhomogeneous equation of motion for sinusoidal force $F(t)=F\cos(\Omega t)$. Although we can not get an analytical solution, the qualitative behavior of the amplitude of chaotic solution $A\propto F^{1/3}$ continuously grows with the increase of $F$, which means no saturation exists in classical Duffing oscillator. Instead of numerical analysis from T. K. Nagy and B. Balachandran\cite{Nagy} we can demonstrate this result qualitatively by assuming external force to take Jacobi Elliptic function form $F(t)=F{\rm cn}(\Omega t, K^2)$, and get a closed form solution from inhomogenous Duffing equation. Jacobi Elliptic function is a generalization of cosine function when taking $K=0$: ${\rm cn}(u, 0)=\cos u$. The closed form solution $x=A\,{\rm cn\,}(\Omega t, K^2)$ with the amplitude $A$ satisfies
\begin{eqnarray}\label{9.5}
A={\frac{2^{\frac{1}{6}}m^{\frac{1}{2}}}{\gamma^{\frac{1}{2}}|2K^2-1|^{\frac{1}{3}}}}
\big|\Delta_-^{\frac{1}{3}}
+
\Delta_+^{\frac{1}{3}}\big|
\end{eqnarray}
where $\Delta_{\pm}=\sqrt{\frac{\gamma}{2}}\frac{F}{m}\pm\sqrt{\frac{\gamma}{2}\left(\frac{F}{m}\right)^2+\frac{4}{27}\frac{\omega^6K^2}{2K^2-1}}$. By taking large $F$ limit, i.e., $F\gg m\sqrt{\frac{8}{27\gamma}\frac{\omega^6K^2}{|2K^2-1|}}$ the amplitude $A\approx \frac{2^{\frac{1}{3}}m^{\frac{1}{6}}}{\gamma^{\frac{1}{3}}|2K^2-1|^{\frac{1}{3}}}F^{\frac{1}{3}}$ increases to infinity as well, implying that saturation cannot exist in classical Duffing oscillator. One should be careful that a closed form solution Eq.(\ref{9.5}) only exists when the external driving force frequency $\Omega$ and $K$ satisfies very special conditions $\Omega=\frac{1}{2^{\frac{1}{3}}(2K^3-K)^{\frac{1}{3}}}\left(\Delta_-^{\frac{1}{3}}+\Delta_+^{\frac{1}{3}}\right)$, $K=\frac{1}{2^{\frac{2}{3}}\Omega}\left(\Delta_-^{\frac{1}{3}}+\Delta_+^{\frac{1}{3}}\right)$ and that the initial conditions $x(t=0)=\dot{x}(t=0)=0$; otherwise a chaotic behavior for $x(t)$ rises again, but the amplitude behavior $A\sim F^{1/3}$ still holds\cite{Nagy}.

Second, we consider quantum infinite-level Hamiltonian ${H}_0$ with increasing enegy spacing $(E_{n+1}-E_n)>(E_n-E_{n-1})$ for $n=1,2,3,...\infty$. Suppose the input phonon is monochromatic with the frequency $\omega$ exactly matches a certain pair of levels $E_m-E_n=\hbar\omega$. The $n, m$-th levels get highly resonant with the increase of phonon amplitude, eventually their populations are equally distributed by phonon emission absorption process: $|a_n|^2=|a_m|^2=1/2$, while any other eigenstate wavefunction are supressed to 0 because the total wavefunction is normalized $\sum_i |a_i|^2=1$. There might be more than one level pair in resonance with $\omega$, however since the level spacing $E_{n+1}-E_n$ is monotonically increasing, for an arbitrary input frequency $\omega$, we can always find a quantum number $N$, such that $E_{N+1}-E_N>\hbar\omega$. Only finite level pairs $n, m < N$ could in resonance with $\omega$, while other levels' wavefunction are supressed to 0 if external phonon amplitude increases to infinity. In experiment ultrasonic phonon frequency usually has a narrow bandwidth $\omega\pm  \Delta\omega$. We can still find a quantum number $N$, such that $E_{N+1}-E_N>\hbar(\omega+\Delta\omega)$ and a similar analysis will be carried out. The above argument also applies to quantum 1-D Duffing oscillator: $
\hat{{H}}(t)=-\frac{\hbar^2}{2m}\partial_x^2+\frac{1}{2}m\omega^2 x^2 +\frac{1}{4}\gamma x^4+\sum_{ij}e_{ij}(\vec x, t)\hat{T}_{ij}
$.

It seems classical and quantum Duffing oscillators contradict to each other, however, we should note in classical problem $\hbar$ is set to be 0. By estimating eigenvalues for a more general nonlinear quantum Hamiltonian we get the quantum number dependence roughly as follows, 
\begin{eqnarray}\label{10}
 & {} & \left(-\frac{\hbar^2}{2m}\partial_x^2+\frac{1}{2}m\omega x^2+\frac{1}{n_0}\gamma x^{n_0}\right)\psi=E\psi\nonumber \\
 & {} & E_n\approx \left(n+\frac{1}{2}\right)\hbar\omega \quad\quad\quad\quad\quad\quad\quad\,\,{\rm for}\,\, n<n_c\nonumber \\
 & {} & E_n\approx \frac{n_0+2}{2n_0}\left(\frac{n^2\hbar^2}{m}\right)^{\frac{n_0}{n_0+2}}\gamma^{\frac{2}{n_0+2}}\quad \,{\rm for}\,\,n>n_c
\end{eqnarray}
where $n_0$ is an integer, $\gamma$ is the positive coefficient for $x^{n_0}$ potential; $n_c=\left(\frac{2n_0}{n_0+2}\right)^{\frac{n_0+2}{n_0-2}}m^{\frac{n_0}{n_0-2}}\omega^{\frac{n_0+2}{n_0-2}}{ \gamma^{-\frac{2}{n_0-2}}}\hbar^{-1}$ is the critical quantum number when nonlinear porential $\gamma x^{n_0}$ starts to be significant. For the special case $n_0=4$ we get quantum Duffing oscillator. If we require that for arbitrary input frequency $\omega$, we can find quantum number $N$ such that the $N$-th energy spacing is greater than $\hbar\omega$, we have
\begin{eqnarray}\label{11}
E_{N+1}-E_N & \approx & \left(\frac{\hbar^2}{m\gamma}\right)^{\frac{n_0}{n_0+2}}N^{\frac{n_0-2}{n_0+2}}\gamma>\hbar\omega\nonumber \\
 & \Rightarrow & N>\omega^{\frac{n_0+2}{n_0-2}}m^{\frac{n_0}{n_0-2}}\gamma^{-\frac{2}{n_0-2}}\hbar^{-1}\nonumber \\
 & \Rightarrow & \lim_{\hbar\to 0}N\to \infty
\end{eqnarray}
In quantum nonlinear oscillator the critical quantum number $N$ is finite; in classical nonlinear oscillator we set $\hbar\to 0$, one can never find a finite quantum number $N$, so that $E_{N+1}-E_N>\hbar\omega$. There are infinite pairs of levels in resonance with external phonons, the energy is unbound for quasi classical nonlinear oscillator. We conclude that saturation is an effect from quantum physics.

\section{Conclusion}
In this paper we generalize TTLS to infinite-level-system without specifying any assumption other than randomness and increasing nonlinearity. By expanding amorphous solid Hamiltonian in orders of strain field $e_{ij}$ in long wavelength limit, the zero-th order $\hat{{H}}_0^{(s)}$ is infinite-level-system; the first order $e_{ij}\hat{T}_{ij}$ is the generalization of TTLS-phonon coupling.

Phonon echo can be proved in infinite-level-system with randomness and nonlinearity. After two coherent pulses being applied the energy levels carry their phases. Before $t<2\tau$ these energy levels' phases are incoherent because of randomness over space and nonlinearity over energy level quantum number. At $t=2\tau$ they become coherent again and emit phonons simultaneously, because as long as one of the level pairs $m, n$ spontaneously emit phonons with phase $(\omega_m-\omega_n)(t-2\tau)|_{t=2\tau}=0$, other level pairs stimulatedly emit phonons with identical frequency $\omega$ and phase $0$. Hence for an infinite-level-system without randomness over space $\vec x$ we cannot find phonon echo effect, because level pairs carrying phase memory are always coherent for $t>5t_0$. They stimulatedly emitting phonons right after each pulse is applied. An infinite-level-system without nonlinearity cannot show phonon echo as well, because there are infinite pairs of levels in resonance with external pulse. These pairs trigger each other and simultaneously emit phonons.

The realization of phonon echo does not require infinite-level-system to have increasing nonlinearity while saturation phenomena does. With the increase of external field, infinite-level-system can not get saturated if the level spacing is equal or decreasing. However one can not prove saturation effect in a classical system, for the fact that $\hbar\to 0$ in classical limit. A detailed discussion was carried out by classical 1-D Duffing oscillator in which the trajectory of displacement is chaotic and the amplitude increases like $\sim F^{1/3}$ with the increase of driving force $F$. Increasingly nonlinear quantum oscillator displays saturation for the reason that for arbitrary input frequency $\omega$ one can always find a finite quantum number $N$ so that the energy spacing $E_{N+1}-E_n>\hbar\omega$. Below $N$ a finite number of level pairs resonantly absorb energy so the system entire energy gets capped.

\section{Acknowledgement}
D. Z. would like to thank Alfred W. Hubler, Dervis C. Vural and Sidney R. Nagel for useful discussions.

\appendix
\widetext

\endwidetext

\end{document}